\documentclass[pre,twocolumn,preprintnumbers]{revtex4}

\usepackage{graphicx}
\usepackage{color}

\def\Dth{\Delta\theta}
\def\Dp{\Delta p}
\def\zb{\bar z}
\def\Imean{\bar I}
\def\bmean{\bar b}

\begin{document}
\title{Out-of-equilibrium mean-field dynamics of a model for wave-particle interaction}

\author{P. de Buyl}
\affiliation{Center for Nonlinear Phenomena and Complex Systems \\
Universit{\'e} Libre de Bruxelles (U.L.B.), Code Postal 231, Campus Plaine, B-1050 Brussels, Belgium }
\author{D. Fanelli}
\affiliation{Dipartimento di Energetica ``Sergio Stecco'', Universit{\'a} di Firenze, via S. Marta 3, 50139 Firenze, Italia}
\author{R. Bachelard}
\affiliation{Synchrotron SOLEIL, L'Orme des Merisiers, Saint-Aubin, F-91192 Gif-Sur-Yvette Cedex, France}
\author{G. De Ninno}
\affiliation{Phys. Dept., Nova Gorica Univ., Nova Gorica (Slovenia)\\
Sincrotrone Trieste, 34012 Trieste, Italy}

\date{\today}
\begin{abstract}
The out-of-equilibrium mean-field dynamics of a model for wave-particle interaction is investigated. Such a model can be regarded as a general formulation for all those applications where the complex interplay between particles and fields is known to be central, e.g., electrostatic instabilities in plasma physics, particle acceleration and free-electron lasers (FELs). The latter case is here assumed as a paradigmatic example. A transition separating different macroscopic regimes is numerically identified and interpreted by making use of the so-called violent relaxation theory.
In the context of free-electron lasers, such a theory is showed to be effective in predicting the saturated regime for energies below the transition.
The transition is explained as a dynamical switch between two metastable regimes, and is related to the properties of a stationary point of an entropic functional.

\end{abstract}

\maketitle

\section{Introduction}

Mean-field models have been widely studied as paradigmatic representatives of the important class of systems subject to 
long-range coupling. In the simplest scenario, $N$ particles are made to interact in one dimension, subject to a varying field which is
self-consistently sensitive to the individual trajectories. A global 
network of connections is hence driving the dynamics of every constituting element, as it 
certainly happens for more realistic settings where e.g. gravity or unscreened 
Coulomb interactions are at play. 

The dynamics of mean-field models displays intriguing features.  Particles may be trapped in intermediate (out-of-equilibrium) states, whose duration diverges with the number of constitutive elements, and which substantially differ from the corresponding thermodynamic equilibrium configuration. 
These metastable states are often termed in the literature Quasi-Stationary States,
hereafter QSS, and bear an extraordinary conceptual importance as they potentially corresponds to the solely 
experimentally accessible regimes, for a wide range of applications ranging from celestial mechanics to plasma physics.

Interestingly, the evolution of the QSS is intimately governed by    
the discreteness of the medium being investigated. More specifically, QSS are believed 
to correspond to stationary stable solution of a Vlasov model, invoked as the continuous analogue 
of the discrete $N$-particles dynamics. 

Within this context, the Hamiltonian Mean Field model (HMF) \cite{antoni_ruffo_clustering} has often been referred to as the benchmark model for elaborating onto the 
QSS emergence. This is a one-dimensional Hamiltonian describing the evolution of $N$ rotors coupled via a mean-field  cosinus-like 
potential.  The QSS in the HMF setting have been explained by resorting to a maximum entropy principle, 
pioneered by Lynden-Bell in astrophysical context, and fully justified from first principles \cite{lynden_bell_1967}. Here, the supposedly relevant Vlasov picture 
enters the description as a Fermionic contribution to an entropy functional. Besides,  
out-of-equilibrium phase transitions are predicted to occur, separating between distinct macroscopic QSSs. 

The Lynden-Bell protocol, also termed violent relaxation theory, was also argued to apply to other mean-field models and, indeed,
it showed effective in predicting the saturate intensity of a free-electron laser (FEL). FELs are lasing devices consisting of a 
relativistic beam of charged particles, interacting with a co-propagating electromagnetic wave. The interaction is assisted by the static and periodic magnetic field generated by an undulator. FEL’s admit a mean-field description in term of the so--called Colson-Bonifacio model\cite{colson_1976,bonifacio_1987}, which captures the essence of the collective wave-particle dynamics. However, no detailed study has been carried out for the FEL case aiming at unravelling the possible existence of  
out-of-equilibrium transition of the type mentioned above. Are these transitions ubiquitous in mean field dynamics and, in this case, 
can we provide a consistent intrepretative framework for their emergence? This paper is dedicated to answering such questions, and  makes 
reference to the specific FEL setting. We also stress that the Colson-Bonifacio model of FEL dynamics can be regarded as a general formulation
for all those applications where the complex interplay between particles and fields is well known to be central, 
e.g. electrostatic instabilities in plasma physics\cite{elskens_escande_book}.  As a closing remark, we notice that, on the practical implication side, 
by disposing of reliable predictive tools on the system evolution, one can aim at guiding the system towards different experimental regimes.

\section{The FEL model}

The Colson-Bonifacio model for the FEL dynamics describes the coupled evolution of the electrons with a co-propagating wave. The equations read:
\begin{equation}
  \left\{\begin{array}{l l l}
      \frac{d\theta_j}{d\bar{z}} &=& p_j, \cr
      & &\cr
      \frac{dp_j}{d\bar{z}} &=& - \mathbf{A} e^{i\theta_j} - \mathbf{A^{\!\ast}} e^{-i\theta_j}, \cr
      & &\cr
      \frac{d\mathbf{A}}{d\bar{z}} &=& \frac{1}{N} \sum_j e^{-i\theta_j},
    \end{array}\right.
    \label{bonifacio}
\end{equation}
where $\theta_j$ stands for the particle phase with respect to that of the optical wave, $p_j$ being its conjugate normalized momentum.
The complex quantity $\mathbf{A} = A_x + i A_y$ represents the transverse field and $N$ the number of electrons composing the electron bunch~\footnote{We assume here that electrons are perfectly resonant with the ponderomotive field generated by the optical wave and by the undulator.}. 

Here $\zb$ labels the longitudinal position along the undulator, and it effectively plays the role of time.
The intensity of the laser field is $I=A_x^2+A_y^2$. As it can be seen from the last of Eqs. (\ref{bonifacio}), the bunching term, $b=\frac{1}{N} \sum_j e^{-i\theta_j}$, is the source of wave amplification.  The bunching quantifies the degree of localization of the electrons in the generalized space of their associated phases.
The above discrete system of equations admits a Hamiltonian formulation to which we shall make reference as to
the N-body model. In the $N \rightarrow \infty$ limit, system  (\ref{bonifacio}) converges to the following Vlasov-wave
set of equations \cite{barre_et_al_pre_2004}: 
\begin{eqnarray}
  \label{eq:vlasovwave}
  \frac{\partial f}{\partial \zb} &=& -p \frac{\partial f}{\partial \theta} + 2 (A_x\cos\theta-A_y\sin\theta) \frac{\partial f}{\partial p},\cr
  \frac{\partial A_x}{\partial \bar z} &=& \int d\theta dp\ f \cos\theta, \cr
  \frac{\partial A_y}{\partial \bar z} &=& -\int d\theta dp\ f \sin\theta. 
\end{eqnarray}

Eqs. (\ref{eq:vlasovwave}) can be simulated numerically, thus allowing us to monitor the evolution of the  
phase space distribution function $f(\theta,p)$ along the $\zb$ axis. In our implementation, we adopt the 
semi-Lagrangian method \cite{sonnendrucker_et_al_jcp_1998}, associated with a cubic spline interpolation 
\cite{NR_in_C}. Results of the numerical integration are also checked versus $N$-body simulations and shown to return a 
perfect matching on relatively short time scale, for large enough values of $N$. On longer times, finite--$N$ corrections do matter.  
The discrete system is in turn sensitive to intrinsic granularity effects, stemming from the intimate finiteness of the simulated medium, 
and progressively migrate from the Vlasov state towards the deputed equilibrium configuration. When increasing its size, 
the system spends progressively more time in the Vlasov-like, out-of-equilibrium regime.  
Formally, in the $N \rightarrow \infty$ limit, it never reaches equilibrium, being permanently trapped in the QSS.

As previously anticipated, our study is hence ultimately concerned with the emergence of QSS’s,
in a context where particles and waves evolve self-consistently. We shall be in particular interested in elucidating the occurrence of out-of-equilibrium 
phase transitions via dedicated numerical simulations, and substantiate our claims analytically. 
In doing so, we will virtually extend the conclusion of \cite{AntoniazziYamaguchiPRL} to a broad spectrum of potentially relevant applications,  
beyond the specific case under inspection. Among others, it is again worth mentioning plasma physics: A formulation equivalent to model 
(\ref{eq:vlasovwave}) is in fact often invoked, when studying the collective effects of beam-plasma dynamics \cite{elskens_escande_book}.

\section{On the initial conditions and their subsequent dynamical evolution}
\label{sec:ICandevo}

Let us turn to discussing our results, as obtained via numerical integration of (\ref{eq:vlasovwave}). In order to make contact with the investigations reported in \cite{AntoniazziYamaguchiPRL}, we shall employ in the following a two-dimensional water--bag initial condition in phase space, which can be seen 
as a rough approximation of a smooth Gaussian profile. A (rectangular) water--bag is formally parametrized by two quantities, namely the semi-width of the spanned 
interval in phase, $\Dth$, and its homologue in the momentum direction, $\Dp$.  
The corresponding expression for $f$ can be cast in the form (see also Fig. \ref{fig:lbwb} top-left):
\begin{equation}
  \label{eq:wbic}
  f(\theta,p) = \left\{\begin{array}{l l}
      f_0 & \mbox{if } |p|\leq \Delta p,\cr
          & \mbox{\ \ \ } |\theta|\leq \Delta\theta, \cr
      0 & \mbox{otherwise.}
  \end{array}\right.
\end{equation}
The initial conditions can be also characterized by defining 
\begin{equation}
\label{eq:bunch}
\left\{\begin{array}{r l}
  b_0= & \frac{\sin\Dth}{\Dth} , \cr
  \epsilon = &\frac{\Dp^2}{6},
\end{array}\right.
\end{equation}
where $b_0$ is the initial bunching, and $\epsilon$ the initial average kinetic energy per particle. Notice that we thus access all possible values of the bunching $b_0 \in [0,1]$ by properly tuning $\Dth$, and all positive energies 
$\epsilon$ by varying $\Dp$. We here limit our discussion 
to the case of vanishing initial optical field, $I_0 \simeq 0$, the relevant parameters' space being therefore solely bound to the plan ($b_0$, $\epsilon$). 
The initial condition here selected is hence solely controlled by these two parameters. In other words, we are specializing on a given bidimensional subset, and deliberately ignore the third, in principle available, direction of the reference parameter space. Quantifying the role of such an additional degree of freedom ultimately amounts to investigate the so-called seeded configuration \cite{yu,deni} and will be the subject of future work.

\begin{figure}[h]
  \centering
  \includegraphics[width=3.3in]{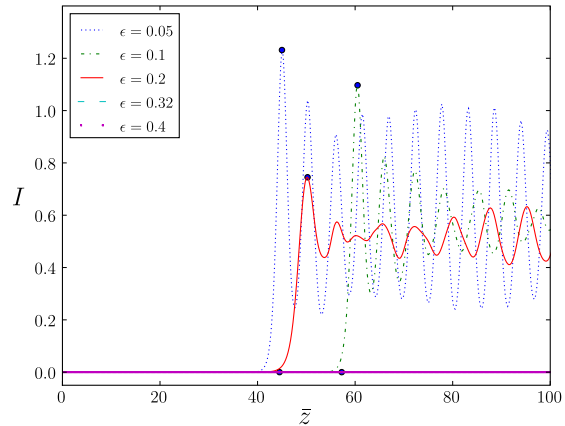}
  \caption{(Color online) As a result of the numerical integration of Eqs. (\ref{eq:vlasovwave}), the evolution of $I$ as a function of $\zb$ is reported for different choices of the energy 
  $\epsilon$, ($\epsilon = 0.05,0.1,0.2,0.3,0.4$); $b_0=0$. Symbols pinpoint the   
  position of the first peak in the intensity time series, thus returning an indication on the 
  saturation time. Notice that for $\epsilon > \epsilon_c=0.315$, 
  the peak is found for $I \ll 1$ : the corresponding initial conditions are hence stable, and no instability develops. }
  \label{fig:b0_0_wb}
\end{figure}

Let us start by discussing the simplest scenario, where the initial beam of particles is uniformly distributed over $[-\pi;\pi]$. From a physical point of view, this amounts to specialize to the case of Self-Amplified Spontaneous Emission (SASE, \cite{pelle1, desy}), where $b_0=0$ and no seed is applied externally. Such a choice was also considered by  
Barr{\'e} {\it et al.} \cite{barre_et_al_pre_2004} and Curbis {\it et al.} \cite{curbis_et_al_eur_phys_j_b_2007},  
where the dependence of the system evolution on the energy was numerically monitored, within the  
$N$-body discrete viewpoint. Interestingly, $b_0=0$ is a stationary solution of the Vlasov system: a local perturbative calculation can hence be straightforwardly 
implemented so to investigate its inherent stability \cite{bonifacio_casagrande_nuovo_cimento_1990}; the calculations are detailed in appendix \ref{sec:stabilitywb}. For $\epsilon < 0.315$, an instability occurs: both 
the wave intensity and the bunching factor rapidly grow, before relaxing towards an oscillating plateau. The average value of $I$ reached in the oscillating regime is called the saturated intensity $\Imean$. This behaviour is displayed in Fig. \ref{fig:b0_0_wb}, where the simulations with $\epsilon > 0.315$ do not show an amplification of $I$. 
This is a well--known property, indeed correctly reproduced by our numerical simulations, and which first signals the existence of phase 
transitions, of the type depicted in \cite{AntoniazziYamaguchiPRL}.
%Let us point here that this critical energy is linked to the stability of our homegeneous initial condition and that when we will consider non-homogeneous ones, the initial behaviour will be non-stationary and a non-linear evolution will follow.
To further corroborate our guess on the $b_0=0$ behaviour, we turn to measuring the saturated intensity $\Imean$ as function of the energy $\epsilon$, where $\Imean$ stands for the mean of $I$ after saturation. It is here computed during four oscillations of $I$.

\begin{figure}[h]
  \centering
  \includegraphics[width=3.3in]{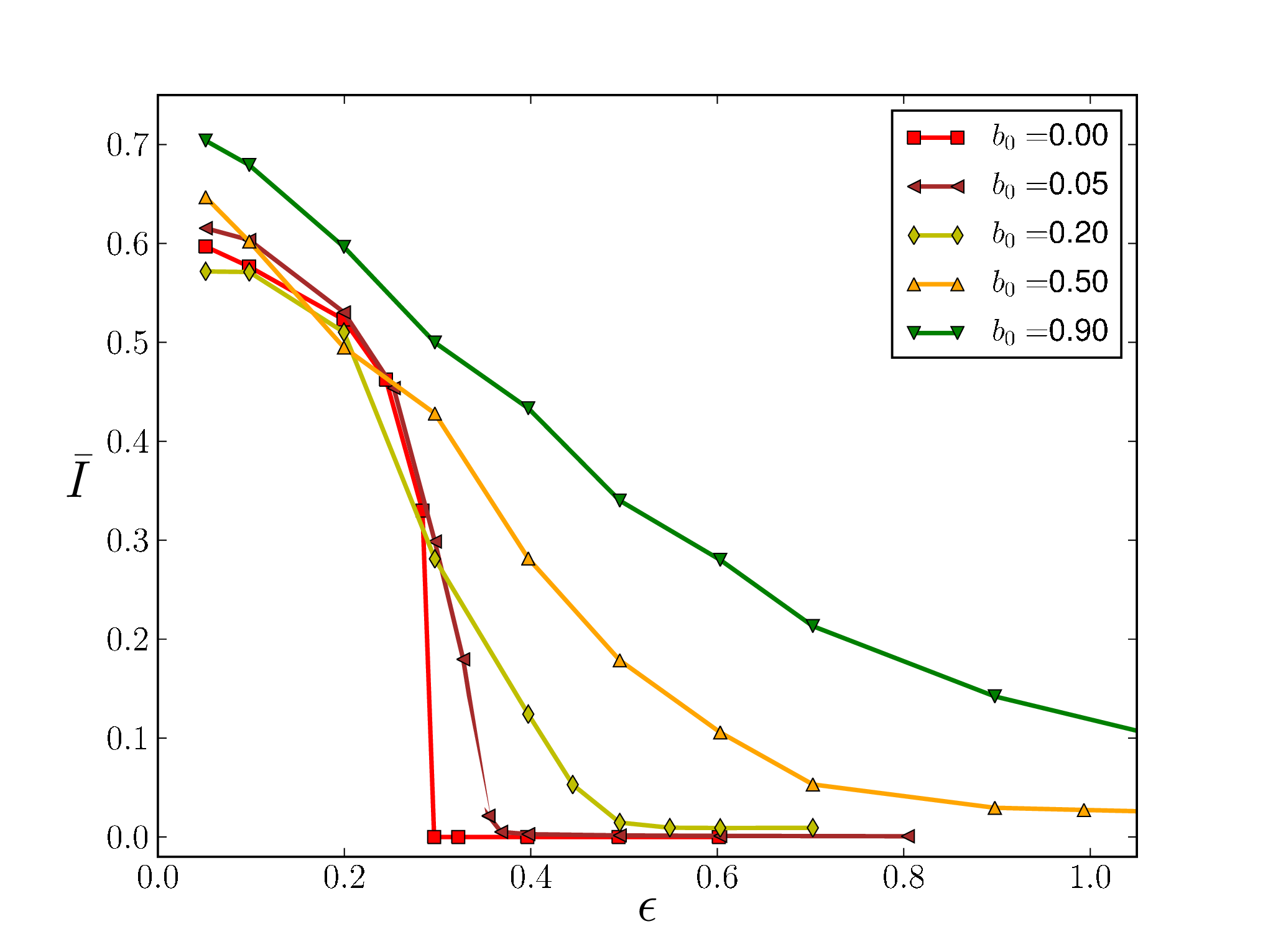}
  \caption{(Color online) Saturated intensity, $\bar{I}$, vs. $\epsilon$ different choices of the initial bunching  $b_0 = 0.0,\ 0.05,\ 0.20,\ 0.50\ \mbox{and } 0.90$}
  \label{fig:b0_all_I_of_e}
\end{figure}

As shown in Fig.  \ref{fig:b0_all_I_of_e}, $\Imean$ rapidly shrinks, 
when increasing the energy $\epsilon$, until a critical value is reached where a sudden transition to $\Imean \simeq 0$ is observed, bearing 
the  characteristic of a first order phase transition. This is a further point of contact with the analysis carried on in \cite{AntoniazziYamaguchiPRL}
for the HMF toy model.

\begin{figure}[h]
  \centering
  \includegraphics[width=3.3in]{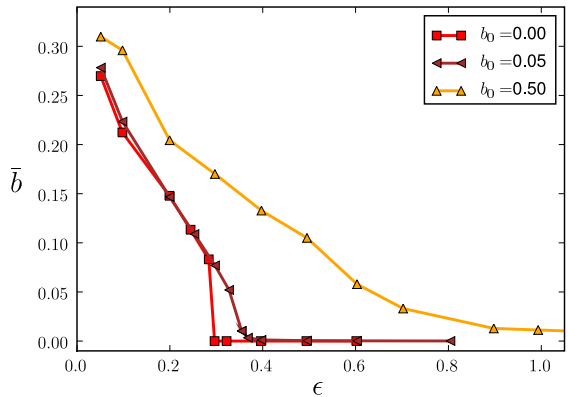}
  \caption{(Color online) Same as Fig. \ref{fig:b0_all_I_of_e}, for the saturated bunching, $\bar b$.}
  \label{fig:b0_all_b_of_e}
\end{figure}

Motivated by these findings, and to push the analogy with the HMF setting, we consider bunched initial distributions. From a physical point of view, this choice is relevant to the case of FEL’s working in the so-called harmonic generation regime \cite{yu, deni}.
Particles' positions are here initially assigned so to  uniformly span a limited portion of the allowed support, symmetric with respect to the origin, controlling the associated bunching via Eq. (\ref{eq:bunch}). 
The inhomogeneous ($b_0 \ne 0$) distribution in phase space is by nature non--stationary. Vlasov dynamics can however smooth it to a homogeneous, 
($b = 0$, $I = 0$) possibly non water--bag, distribution or evolve to a bunched situation. The saturated mean-field average intensity 
$\Imean$ vs. the energy parameter $\epsilon$ is represented in Fig. \ref{fig:b0_all_I_of_e}, 
showing the newly collected data for different values of $b_0>0$ to the reference profile relative to $b_0=0$. In all cases the intensity is shown to decrease, 
as the energy increases. Importantly, for small values of $b_0$, an abrupt transition is observed, which can be naively interpreted as of the
first--order type. For larger values of $b_0$, the observed transition becomes smoother, such as for a second--order one. A substantially 
identical scenario holds for the bunching, which evolve towards an asymptotic plateau $\bmean$, also sensitive to the $\epsilon$ and 
$b_0$ parameters, see Fig. \ref{fig:b0_all_b_of_e}.  This scenario points towards a unifying picture on the 
emergence of out-of-equilibrium phase transitions within the considered class of mean-field Hamiltonian model. As previously anticipated, 
the Lynden-Bell theory of violent relaxation was successfully applied to the HMF problem, allowing one to gain a comprehensive 
understanding on the out-of-equilibrium phase transition issue, including a rather accurate characterization of the associated transition order. In the 
following section we set down to apply the Lynden-Bell argument to the present case, benchmarking the theory to numerical 
experiments.

\begin{figure}[h]
  \centering
  \includegraphics[width=3.3in]{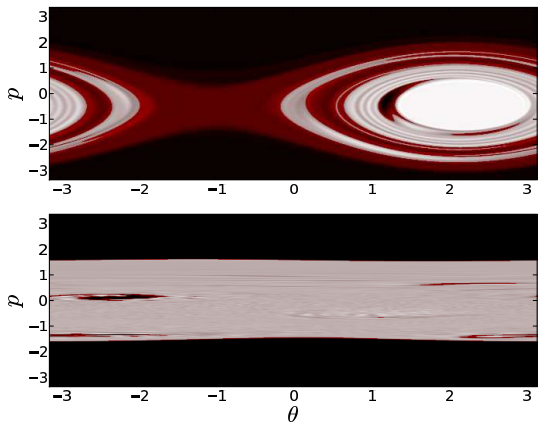}
  \caption{(Color online) Phase space density $f(\theta,p)$ for $b_0=0.05$ and $\epsilon = 0.2 \mbox{ (top) and } 0.4 \mbox{ (bottom)}$.}
  \label{fig:PhaseSpace}
\end{figure}

Before ending this section, we briefly discuss the phase--space structures resulting from 
the Vlasov-based simulations. Two phase--space portraits are enclosed in Fig. \ref{fig:PhaseSpace}, and refer to different values 
of the energy $\epsilon$, respectively below (upper panel) and above (lower panel) the critical transition energy relative
to the selected (fixed) $b_0$ amount. When the system evolves towards a state at $\Imean \ne 0$, then $f(\theta,p)$  shows  a  
large resonance. At variance, in the opposite regime, a hole-resonance dipole structure is observed, see also
\cite{AntoniazziProceedings}.  This observation seems to suggest that the out-of-equilibrium phase transition materializes via a bifurcation of invariant structures, 
an observation that was recently made for the HMF model \cite{BachelardFanelliPRL}. Results of further investigations on this specific topic   
will be presented in a separate contribution \cite{BachelardInPreparation}.

\section{On the violent relaxation theory}

In his work on self-gravitating systems, Lynden-Bell suggested \cite{lynden_bell_1967} that the collisionless dynamics governed by the Vlasov
equation tends to maximize a Fermionic entropy. The latter is obtained from the classical definition, where the counting of the microscopic 
states, compatible with a given macroscopic configuration, results from a combinatorial calculation, and is sensitive to the underlying Vlasov dynamics.  The method was successfully employed    
in the study of the HMF model \cite{antoniazzi_et_al_prl_2007,chavanis_epjb_2006,AntoniazziYamaguchiPRL} and also applied to predict the quasi-stationary amplitude of the FEL wave~\cite{barre_et_al_pre_2004,curbis_et_al_eur_phys_j_b_2007}. In these works, however, the analysis just focused on the unstable regime 
($\Imean \ne 0$): no attempt was in fact made to reconcile it, with the high energy homogeneous state, via the phenomenon of out-of-equilibrium phase transitions.  
More recently, Yamaguchi \cite{yamaguchi_pre_2008} used the Lynden-Bell approach to predict the core of the gravitational sheet model,
demonstrating its adequacy within the field for which it was originally conceived. In the following we shall review the main steps of the
derivation of the violent relaxation theory, applied to the FEL setting. Starting from a water--bag, 
the entropy to be maximized \cite{lynden_bell_1967} can be cast in the form \cite{chavanis_sommeria_miller_1996}
\begin{equation}
  \label{eq:LBentro}
  s(\bar f) = - \int dp\ d\theta\ \left[ \frac{\bar f}{f_0}\ln\frac{\bar f}{f_0} + \left(1-\frac{\bar f}{f_0}\right)\ln\left(1-\frac{\bar f}{f_0}\right)  \right],
\end{equation}
where $f_0$  is specified in (\ref{eq:wbic}) and $\bar f$ is the coarse grained distribution function. 
Following Barr{\'e} {\it et al.} \cite{barre_et_al_pre_2004,barre_phd}, maximizing the functional  (\ref{eq:LBentro}), results in the following set of equations
\begin{eqnarray}
  \label{eq:systemofeq}
    f_0 \frac{x}{\sqrt{\beta}} \int d\theta\ \zeta F_0(\zeta x) &=& 1, \cr
    f_0 \frac{x}{\sqrt{\beta}} \int d\theta\ \sin\theta\ \zeta F_0(\zeta x) &=&  A^3, \cr
    f_0 \frac{x}{2 \beta^{1.5}} \int d\theta\ \zeta F_2(\zeta x) &=& \epsilon + \frac{3}{2} A^4,
\end{eqnarray}

where $\zeta  = \exp\left(-2 A \beta \sin\theta \right)$, $F_0(y)=\int_{-\infty}^\infty \frac{e^{-\frac{v^2}{2}} dv}{1+y\ e^{-\frac{v^2}{2}}}$ and 
$F_2(y)=\int_{-\infty}^\infty \frac{v^2 e^{-\frac{v^2}{2}} dv}{1+y\ e^{-\frac{v^2}{2}}}$. Here $\beta$ and $x$ are (rescaled) Lagrange multipliers and
ultimately stem from the conservation of mass, momentum and energy. $A$, $\beta$ and $x$ are calculated by solving  Eqs. (\ref{eq:systemofeq}) 
numerically via a Newton-Raphson method.  The resulting (real) value of $A$ is expected to return an estimate of the laser intensity at (Vlasov) saturation, $\Imean = A^2$, while $f(\theta,p)$ is:
\begin{equation}
  f(\theta,p) = f_0 \frac{1}{1+x\ e^{\beta (p^2/2 + 2 A \sin\theta + A^2p + A^4/2)}}.
\end{equation}

For $A=0$ (namely $\Imean=0$) the optimization problem (\ref{eq:systemofeq}) reduces to:

\begin{equation}
  \label{eq:oneeq}
  x=\sqrt{12 \frac{F_2(x)}{F_0(x)^3} \frac{\Dth}{\pi}}.
\end{equation}

\begin{figure}[h]
  \centering
  \includegraphics[width=3.3in]{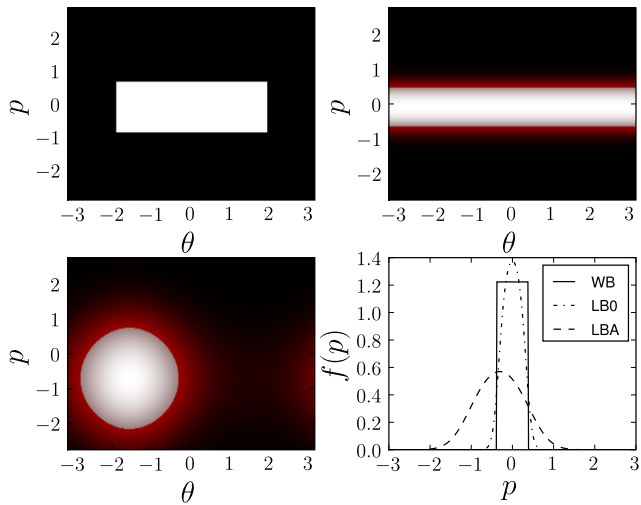}
  \caption{(Color online) Examples of $f(\theta,p)$: (top to bottom, left to right) the waterbag, the LB0 solution, the LBA solution and the velocity distribution function integrated over space for the three solutions.}
  \label{fig:lbwb}
\end{figure}

An $x$ value exists which solves the above equation for any choice of $\Dth$. The homogeneous state is a stationary 
solution of the Lynden-Bell entropy, and so a potentially attractive state of the Vlasov dynamics. 
Additional inhomogeneous solutions ($A \ne 0$, or, equivalently, $\Imean \ne 0$)  might however emerge from investigating the full 
system (\ref{eq:systemofeq}). The homogeneous and inhomogeneous solutions will be referred to as to LB0 and LBA, 
respectively (see Fig. \ref{fig:lbwb}). The forthcoming discussion will focus on how to discriminate between the two, and eventually predict the asymptotic fate of the system. 

\noindent

\section{Interpreting the out-of-equilibrium transition as a dynamical switch between LBA and LB0}
\label{sec:lb0andlba}

\begin{figure}[h]
  \centering
  \includegraphics[width=3.3in]{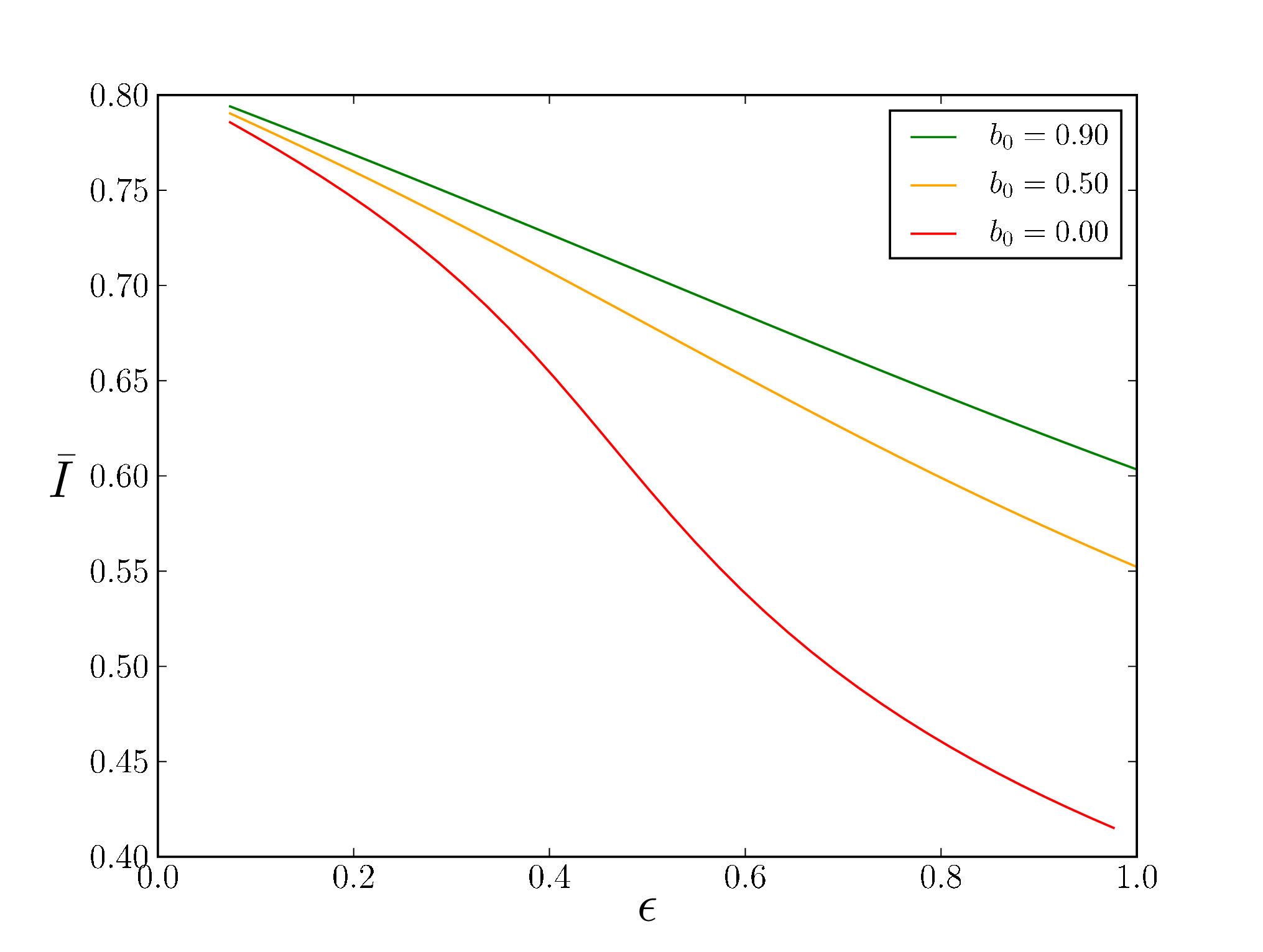}
  \caption{(Color online) The saturated intensity $\Imean$ for the (inhomogeneous) LBA solution of system (\ref{eq:systemofeq}) plotted as 
  function of the energy $\epsilon$. Different curves refer to distinct value of the initial bunching $b_0$ (see legend).
  }
  \label{fig:LBA_A}
\end{figure}

\begin{figure}[h]
  \centering
  \includegraphics[width=3.3in]{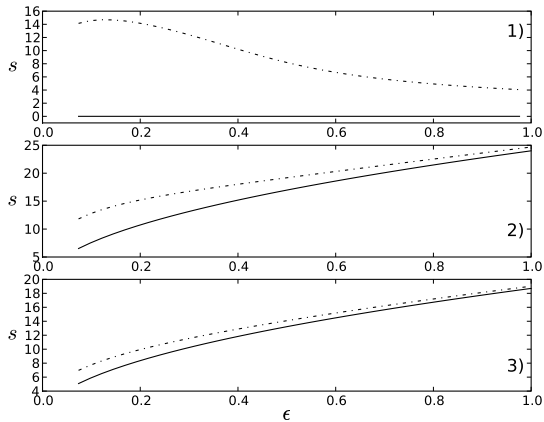}
  \caption{The Lynden-Bell entropy calculated respectively for the LB0 (plain line) and LBA (dash-dotted line) solution. Here 
  $b_0 = 0.00,\ 0.50,\mbox{ and } 0.90$ (resp. panels 1,2 and 3).}
  \label{fig:LB_entropy}
\end{figure}

Let us focus first on the LBA solution. In Fig. \ref{fig:LBA_A}, we report  the value of $\Imean$ as it follows from Eqs.~(\ref{eq:systemofeq}), for different choices of the energy and initial bunching.
The predicted intensity is shown to decrease, when the energy gets larger, but no transition is observed, in contradiction with the 
results of our numerical simulations. As previously stressed, the homogeneous LB0 state is also solution of the optimization problem
(\ref{eq:systemofeq}) and could in principle prevail over the former. To shed light on this issue, we 
calculated the entropy values $S_A$ and $S_0$, associated to LBA and LB0, respectively. 
Results of the computations are shown in Fig. \ref{fig:LB_entropy}, where the dependence on the energy $\epsilon$
is monitored for various choices of $b_0$. Surprisingly, and at odd with what happens for the HMF model \cite{AntoniazziYamaguchiPRL}, 
$S_A$ is always larger than $S_0$. The two curves do not cross each other and the LBA configuration is entropically favoured. Let us note that for higher values of $\epsilon$, $\Imean$ decreases and the two solutions get close to each other, also from the point of view of the entropy, without crossing.

The observed transition could possibly stem from a purely dynamical mechanism, 
and this would justify the discrepancy between the simulation output and the statistical prediction.
More specifically, we here argue that, depending on the selected initial conditions, 
the system explores a local basin of attraction and struggles to find its way to the deputed, 
global maximum of the entropy. To clarify this point, we focus on a single numerical simulation, assuming the system to be initialized in a LB0 state. 
The dynamics can progressively take the system towards the LBA configuration, respecting the maximization of the Lynden-Bell entropy.  
The opposite is not possible, and a simulation started in the LBA state will certainly not evolve to the LB0. However, dynamical effects might be also 
at play and interfere with ideal situation here schematized, by virtually blocking the system in the neighborhood of an initially assigned 
LB0 conformation. Is this the correct scenario? And how can one explain the observed transition when starting from  waterbag initial conditions? These
issues are addressed in the following, where the stability of LB0 and LBA is investigated via direct Vlasov simulations.

\begin{figure}[h]
  \centering
  \includegraphics[width=3.3in]{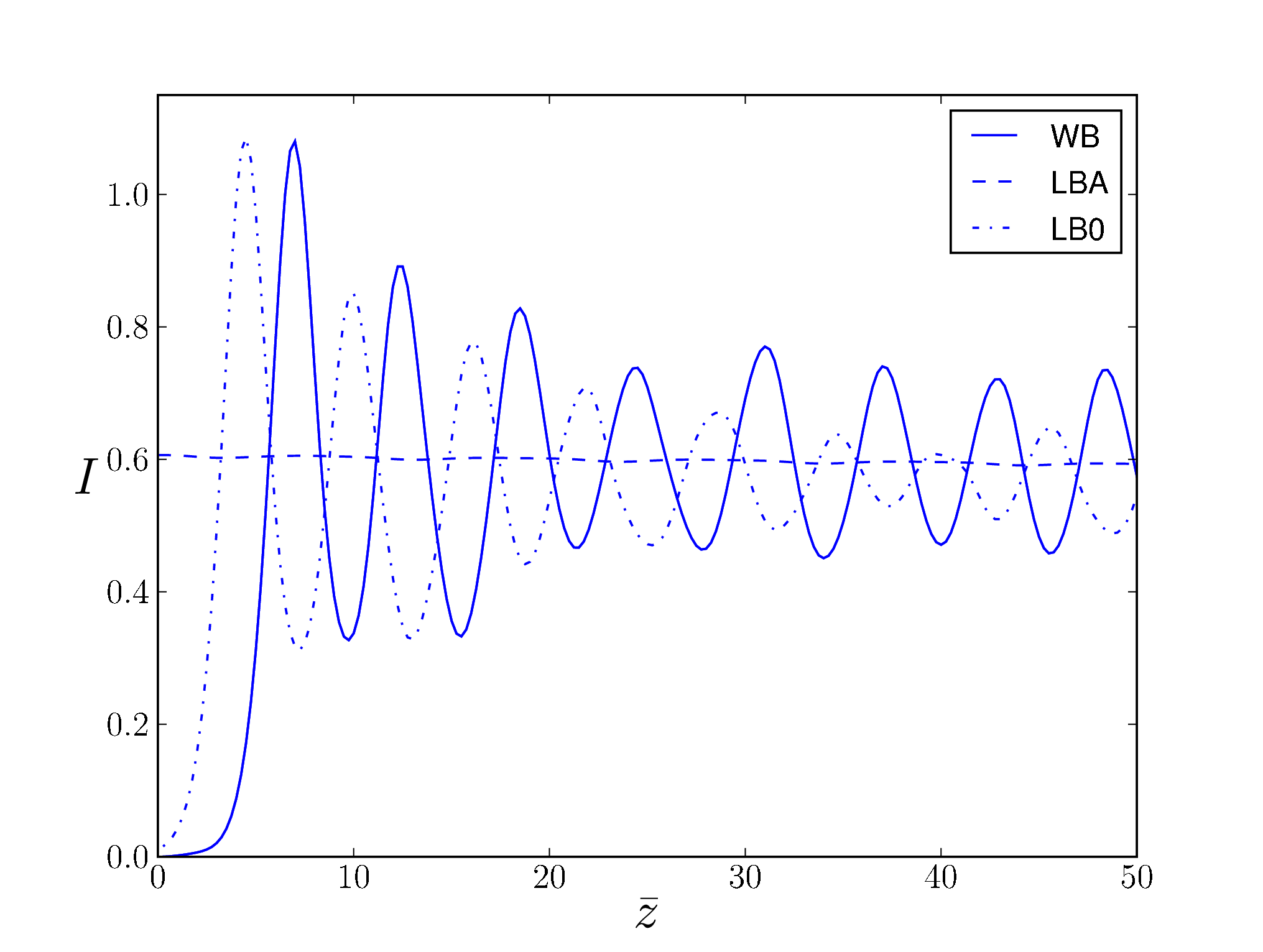}
  \caption{(Color online) The intensity $I$ as a function of time, for three different 
  choices of the initial condition: WB (plain line), LBA (dashed line) and 
  LB0 (dotted line). All conditions refer to $b_0=0.05$,$\epsilon=0.10$}
  \label{fig:FEL_LBA_stable}
\end{figure}

In Fig.  \ref{fig:FEL_LBA_stable} the dynamical evolution of the intensity $I$ is depicted, for three different classes 
of initial conditions, relative to the same choice of $\epsilon$ and $b_0$. The LBA is indeed stable, no deviation from the initial configuration being observed as an effect of the Vlasov dynamics. Conversely, the LB0 condition proves unstable, and the intensity converges towards an oscillating plateau. 
Interestingly, the LB0 and water--bag (WB) evolutions  are qualitatively similar, and, moreover, display the same average 
asymptotic value for the intensity $I$. Even more important, the asymptotic value corresponds to the LBA (maximum entropy) solution.

\begin{figure}[h]
  \centering
  \includegraphics[width=3.3in]{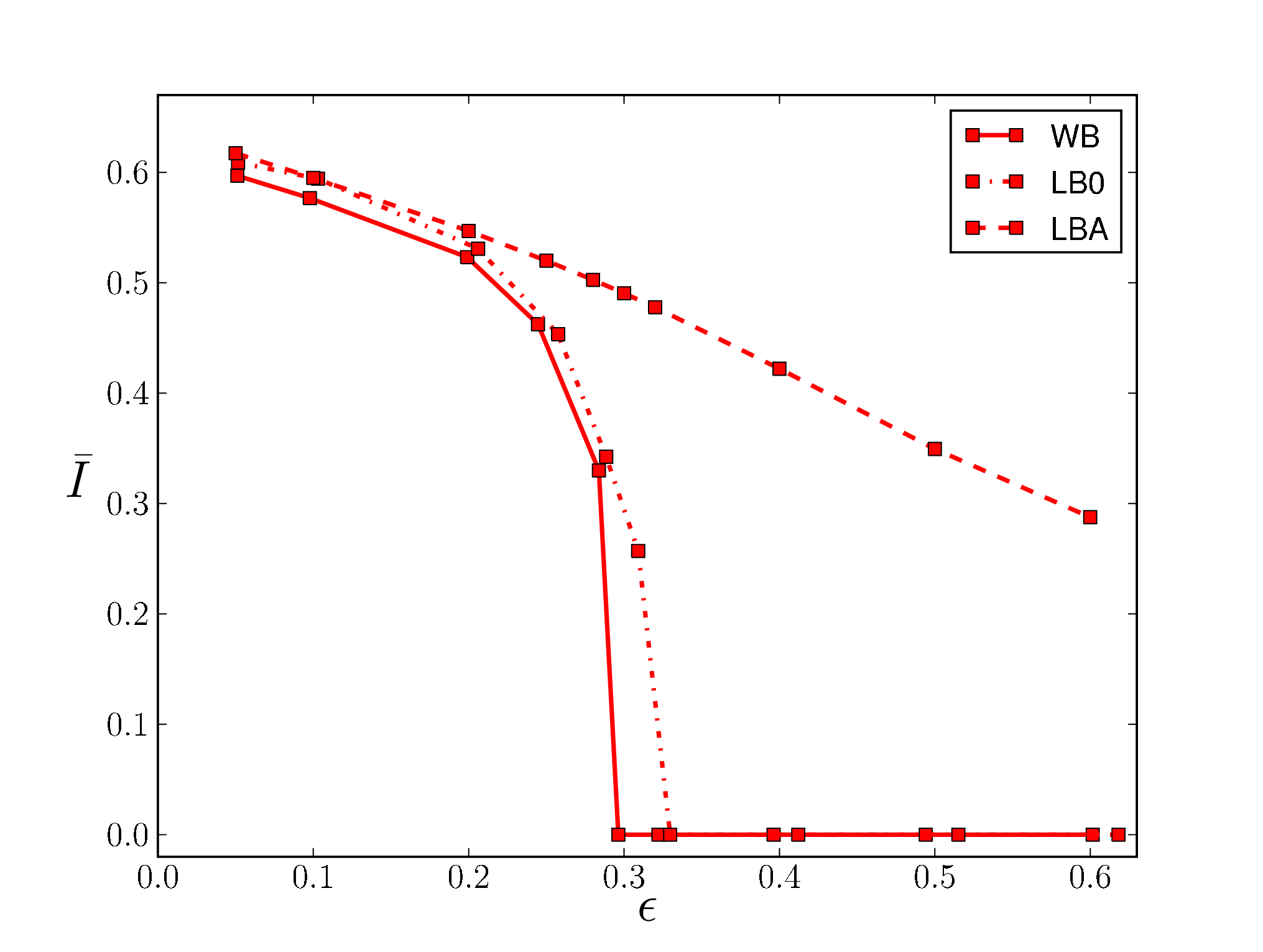}
  \caption{(Color online) The saturated intensity $\Imean$ is plotted as function of the energy $\epsilon$, for $b_0=0$.  WB, LB0 and LBA initial 
  conditions are considered, see legend.}
  \label{fig:FEL_comp_0.00}
\end{figure}

\begin{figure}[h]
  \centering
  \includegraphics[width=3.3in]{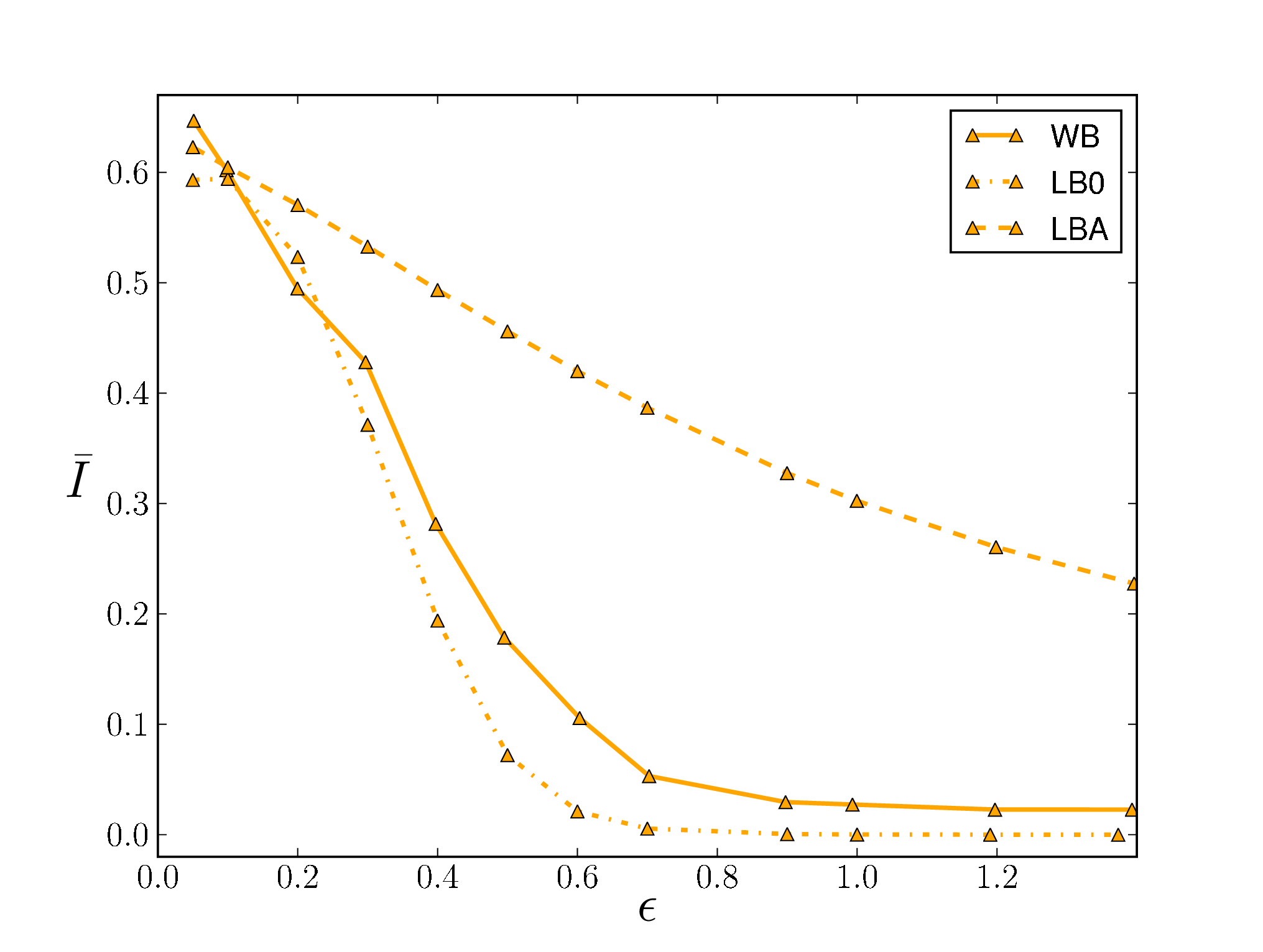}
  \caption{(Color online) The saturated intensity $\Imean$ is plotted as function of the energy $\epsilon$, for $b_0=0.50$.  WB, LB0 and LBA initial 
  conditions are considered, see legend.}
  \label{fig:FEL_comp_0.50}
\end{figure}

To further elucidate the analogies between LB0 and WB, and also clarify the stability issue, we performed a detailed campaign of 
simulations, aimed at generalizing the results of Fig.  \ref{fig:FEL_LBA_stable}. Results of 
the investigations are reported in Figs.  \ref{fig:FEL_comp_0.00} and \ref{fig:FEL_comp_0.50}, where $\Imean$ is represented versus 
$\epsilon$, for two selected $b_0$ amounts. The LBA initial condition is always stable under 
Vlasov dynamics. It is in fact a global maximum of the Lynden-Bell entropy, which, in this respect, proves adequate to describe the system 
at hand. The observed LB0 evolution is by far more complex. For large values of the energy, LB0 is stable. The stability is eventually lost 
when reducing the energy parameter: A transition materializes and the LB0 evolves towards the LBA state. In the vicinity of 
the transition, the LB0 initial condition approaches an asymptotic configuration, which slightly differs from the LBA one, and
possibly results from a balance between opposing dynamical strengths. The WB evolution mimics that of LB0, the two curves returning a pretty close 
correspondence. In practice, during a short transient, an initially bunched WB expands (almost) ballistically, 
the particles being essentially transported by their own initial velocities, so visiting the whole interval $[-\pi;\pi]$. The obtained 
distribution can be approximated by a homogeneous LB0 state (the field has not yet developed, its intensity being effectively negligible), 
which in turn explains the observed correspondence. We shall however emphasize 
that this mechanism applies to relatively small $b_0$ amounts. This fact is testified in Fig.  \ref{fig:FEL_comp_0.50}: 
The agreement between LB0 and WB evolution is shown to worsen, when compared to that of Fig. \ref{fig:FEL_comp_0.00}.
This is understood as follows: Starting from a high degree of bunching, the induced field opposes the natural ballistic contribution, 
by further enhancing the tendency to form a coherent clump of particles. 

In summary, our calculation returns two stationary points of the Lynden-Bell entropy. The first, which we termed LBA, corresponds to 
a inhomogeneous (laser on) configuration and it is a global maximum of the entropy. The second, labelled LB0, is homogeneous (laser off).
For sufficiently large energies, the system can be locally trapped in the vicinity of the LB0. This happens if the system is initiated close enough to a LB0 state, as e.g. in the case of WB with moderate $b_0$ values.
For smaller $\epsilon$, the LB0 loses stability and the system departs towards the entropically favoured LBA state.
%Having detected no additional stationary points of the fermionic entropy, other than LB0 and LBA, we interpret the observed transition as a change of stabiity of LB0 which is a saddle point.
%Of all direction of evolution in $f(\theta,p)$-space, the one giving rise to the observed instability (eventually leading to LBA) becomes stable and the system is trapped near LB0, at least for the duration of our numerical experiments.

Having detected no additional stationary points of the fermionic entropy, other than LB0 and LBA, we interpret LB0 and LBA as a saddle point and a global maximum, respectively.
%While LB0 could in principle also be a global minimum, this hypothesis is invalidated by the fact that there exist (at least) a $f$ with a lower entropy.
%We give such a $f$ which, while not a stationary point of Lynden-Bell's entropy, is a possible stationary solution of system (\ref{eq:vlasovwave}) in the coarse-grained point of view implied by the theory~:
While LB0 could also be in principle a global minimum, this hypothesis is invalidated by the fact that there exists (at least) a function $f$, compatible with the system dynamics, which
yields a lower entropy value.  Consider in fact~:
\begin{equation}
  \label{eq:fLowerS}
  f(\theta,p) = \left\{\begin{array}{l l}
      f_0 \times \frac{\Dth}{\pi} & \mbox{if } |p|\leq \Delta p,\cr
      0 & \mbox{otherwise.}
  \end{array}\right.  
\end{equation}

This is not a stationary point of Lynden-Bell's entropy but represents one of the admissible equilibria of the dynamical system (\ref{eq:vlasovwave}), in its coarse grained perspective as implied by the theory.
The Lynden-Bell entropy associated to Eq. (\ref{eq:fLowerS}) can be straightforwardly computed via Eq. (\ref{eq:LBentro}) and it is found to be always smaller than $S_0$, the value associated to the LB0 configuration, for all $b_0$ and $\epsilon$. Based on the above, and as previously anticipated,  we can exclude the possibility for LB0 to be a global minimum.  Following our deductive reasoning, we hence suggest that LB0 is instead a saddle-point and the observed transition is consequently interpreted to stem from a local modification of LB0 stability properties or morphological characteristics (e.g. width/flatness of the stability basin). A detailed analytical characterization of LB0 stability is at present missing and could eventually help clarifying the underlying scenario.

%Lynden-Bell's entropy associated with Eq. (\ref{eq:fLowerS}) can be computed using Eq. (\ref{eq:LBentro}); it always falls below $S_0$, the value associated with LB0, for all values of $b_0$ and $\epsilon$. We are thus left with LB0 as a saddle point; we suspect a change in its shape (width of the stability basin) or stability at the transition, but analytical knowledge is missing on that point.

\section{Conclusions}

Using the case of a FEL as a paradigmatic example, in this paper we described the out-of-equilibrium dynamics of a mean--field model for wave-particle interaction. Our numerical investigation moves from the Vlasov version of the model, which rigorously applies to 
the continuous limit and is believed to constitute the correct interpretative framework to elaborate on the QSS 
peculiarities. Working within this context, and assuming a specific class of initial conditions, we identified   
a switch between different macroscopic regimes.  Such a transition is ultimately controlled by the nominal 
energy value and by the initial particles bunching, in qualitative agreement with what was previously observed for the HMF model.

The Lynden-Bell violent relaxation theory is developed with reference to the FEL setting to quantitatively substantiate our findings. 
We numerically characterized the stability of the stationary points of the Fermionic entropy functional.
A sudden change in their characteristic is related to the occurrence of the observed transition, 
supporting the adequacy of the violent-relaxation theory to describe the states reached by the dynamical system.

As a final comment, and beside stressing the unifying picture that is here brought forward, we 
emphasize that the transition here predicted can be in principle observed in real devices.
We regard this as a rather important series of experiments 
which could eventually result in a direct proof on the existence of QSS for wave--particle systems.  

\appendix
\section{Stability of the homogeneous waterbag}
\label{sec:stabilitywb}

This Appendix is devoted to reviewing the stability condition 
of system (\ref{eq:vlasovwave}) for an initial homogeneous distribution of the waterbag type. 
The derivation follows from a straightforward linear analysis which can be found for
instance in \cite{elskens_escande_book}. We shall hereafter make reference to the calculation detailed
in \cite{romain_duccio_cnsns_2008}. Let us start by assuming a general 
equilibrium setting where the spatial distribution is homogeneous ($A_x=A_y=0$)
and $f=f_0(p)$, i.e. a generic function of the variable $p$. Then one can
linearize around the equilibrium and eventually derive an explicit 
solution which holds for a relatively short time. To this end we write:

\begin{eqnarray}\label{pertubrsolu}
  &f(\theta,p,t) &=f_0(p)+f_1(\theta,p,t),\cr
  &A_x(t)&=X_1(t)\quad\mbox{and}\quad A_y(t)=Y_1(t)\quad.
\end{eqnarray}
where the quantities labeled with the index $1$ stand for the linear 
perturbation. Introducing in system (2) and retaining the lowest order yields:

\begin{eqnarray}
 (\partial_{\bar{z}}+p\partial_\theta)f_1 -2\eta(X_1 \cos \theta&-Y_1 \sin \theta)&=0\label{deuxsolia}\\
 \int_{-\pi}^{\pi}\!\! d \theta\!\!\int_{-\infty}^{+\infty}\!\! d p\  f_1\cos \theta&
 -\frac{d X_1}{d \bar{z}} &=0\label{deuxsolib}\\
  \int_{-\pi}^{\pi}\!\! d \theta\!\!\int_{-\infty}^{+\infty}\!\! d p\  f_1\sin \theta&
 +\frac{d Y_1}{d \bar{z}} &=0\label{deuxsolic}
 \end{eqnarray}
where we introduced $\eta(p)=\partial_p f_0(p)$. The above linear system admits a
solution in terms of normal modes:

\begin{eqnarray}
f_1(\theta,p,\bar{z})&=&F_1(p)\, e^{i(\theta-\omega \bar{z})}+
F_1^*(p)\, e^{-i(\theta-\omega^* \bar{z})}\label{solf1}\\
X_1(\bar{z})&=&X_1\, e^{-i\omega \bar{z}}+X_1^*\, e^{i\omega^* \bar{z}}\\
Y_1(\bar{z})&=&iY_1 \, e^{-i\omega \bar{z}}-iY_1^*\, e^{i\omega^*
\bar{z}}\quad.\label{solY1}
\end{eqnarray}
where the symbol $*$ refers to the complex conjugate and in general 
$\omega\in \mathbf{C}$. Making use of the above ansatz in the linearized system
of equations returns the following consistency equation:

\begin{equation}\label{dispersionfinal}
\omega =  \int_{-\infty}^{+\infty}\!\! dp
\,\frac{\partial_p f_0}{p-\omega}\quad
\end{equation}
often referred to as to the dispersion relation. To determine whether a given
distribution $f_0(p)$ is stable or unstable, one can solve the above dispersion
relation and estimate the sign of the imaginary part of  $\omega$. Depending
on the sign the field grows exponentially (instability) or 
oscillates indefinitely (stability). If the selected initial condition is
parametrized via an adjustable parameter, one can then calculate the
corresponding theshold value which discriminates between stable and 
unstable regimes. This analytical procedure can be persecuted in simple cases,
as the one addressed in this paper (the waterbag). For more complicated
situations one can resort to the celebrated Nyquist method, first introduced in   
plasma physics \cite{nyquist} (see also \cite{chavanis_delfini_2009}). For the case at hand,
$f_0(p)$ takes the form:

\begin{equation}\label{ic}
f_0(p)= \frac{1}{2 \pi} \frac{1}{2 \Delta p} \left[
\Theta(p+\Delta p) - \Theta(p-\Delta p) 
\right]
\end{equation}
where $\Theta$ stands for the Heaviside function. Inserting (\ref{ic})
into (\ref{dispersionfinal}), carrying out the integral explicitly and looking
for the value of $\Delta p$ which sets the transition between complex and real
$\omega$, leads to $\Delta p \simeq 1.37$ or equivalently 
$\epsilon = (\Delta p)^2 / 6 \simeq 0.315$.

\section*{Acknowledgments}

PdB would like to thank P. Gaspard and N. Goldman for their support. The research of PdB is financially supported by the Belgian Federal Government (Interuniversity Attraction Pole ``Nonlinear systems, stochastic processes, and statistical mechanics'', 2007-2011).

\end{document}